\documentstyle[aps,floats,epsf]{revtex}
\begin{document}
\twocolumn[\hsize\textwidth\columnwidth\hsize\csname 
           @twocolumnfalse\endcsname
\title{Black-hole interiors and strong cosmic censorship}
\author{Eric Poisson \\
         Department of Physics, University of Guelph \\
         Guelph, Ontario, N1G 2W1, Canada}
\date{September 10, 1997}
\maketitle
\begin{abstract}
\widetext
Strong cosmic censorship holds that given suitable initial data 
on a spacelike hypersurface, the laws of general relativity 
should determine, completely and uniquely, the future evolution 
of the spacetime. Here it is argued that while strong cosmic
censorship is enforced for all black holes residing in
asymptotically flat spacetime, it is violated (within the classical 
formulation of general relativity) for some black holes residing in 
non asymptotically flat spacetime. It is suggested that the
semi-classical formulation of general relativity might enforce 
strong cosmic censorship. 
\end{abstract}
\vskip 2pc]

\narrowtext
\section{Introduction}
\label{poi-intro}

The basic question underlying the theoretical study of black-hole 
interiors is ``what is the structure of spacetime inside a realistic 
black hole?''. A lot of progress has been made during the last few years 
toward answering this question, as will be obvious from the other 
contributions to these proceedings. The question I wish to consider 
in this contribution is the following, which is at once much more 
focused and much more fundamental: ``Do the laws of general relativity 
uniquely determine the structure of spacetime inside the black hole, 
given suitable initial data placed at the onset of gravitational 
collapse?''. I will argue that the evidence points to a negative 
answer in the case of the purely {\it classical} laws, but that there is 
hope for a positive answer in the framework of the {\it semi-classical} 
laws. This contribution is based mostly on a 1992 paper co-authored with
Patrick Brady \cite{poi-1} and a 1995 paper co-authored with Draza 
Markovi\'c \cite{poi-2}. However, the point of view expressed here is 
entirely my own, and Patrick and Draza should not be held liable!

\begin{figure}
\begin{center}
\mbox{\epsfxsize=3in \epsffile{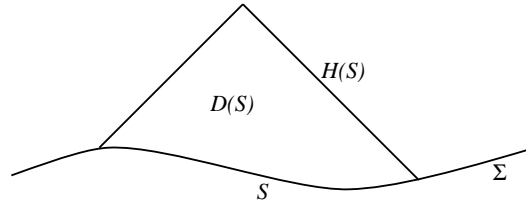}}
\end{center}
\caption{A spacelike hypersurface $\Sigma$ (assumed without an edge) 
contains a set $S$. The evolution of initial data put on $S$ is
determined uniquely only within $D(S)$, the domain of dependence
of $S$. The boundary of this region is $H(S)$, the Cauchy horizon 
of $S$. Strong cosmic censorship holds that as $S$ is made to
coincide with $\Sigma$, then $D(S)$ must coincide with $M$, the entire
spacetime manifold. Then $H(\Sigma) = \emptyset$. In the diagram,
only the future parts of $D(\Sigma)$ and $H(\Sigma)$ are shown.}
\label{poi-fig1}
\end{figure}

The scope of the question should be clarified before attempting
to answer it. What is at stake here is the {\it global} structure 
of spacetime, that is, the full characterization of physical fields, 
including the metric, everywhere and at all times. The well-posedness 
of the initial value problem in general relativity \cite{poi-3} guarantees 
only that given suitable data on an initial surface, the solution to the 
Einstein equations will be unique (up to diffeomorphisms) everywhere 
within the {\it domain of dependence} of the initial surface. Denoting 
this surface by $\Sigma$ and its domain of dependence by $D(\Sigma)$, 
the question facing us is whether $D(\Sigma)$ coincides with $M$, the 
entire spacetime manifold. (See Fig.~\ref{poi-fig1}.) In other words, 
is there a region in the manifold which does not lie inside the domain of 
dependence of the initial surface? If we define the {\it Cauchy horizon} 
$H(\Sigma)$ of the initial surface to be the boundary of its domain of 
dependence, then the question is whether there exists a Cauchy horizon 
at all, and the answer is negative if $D(\Sigma) = M$.

This set of questions is usually grouped under the name {\it strong
cosmic censorship}. The principle of strong cosmic censorship expresses
the basic idea that starting from suitable initial data, general
relativity should be able to predict, unambiguously, the complete
future evolution of spacetime. As we have seen, a more technical 
way of expressing this idea is that the domain of dependence of the 
initial surface should be the entire spacetime manifold, or that the 
Cauchy horizon of the initial surface should be the empty set. A
spacetime which satisfies these properties is said to be {\it globally
hyperbolic}. A much more precise proposal for a strong cosmic censorship 
conjecture can be found in Wald's book \cite{poi-3}. 

A more frequently discussed form of cosmic censorship is the weak
form, which states, roughly, that physically realistic spacetimes
should not contain any globally naked singularities. A singularity 
is globally naked if it can be detected by observers at arbitrarily 
large distances. A singularity hidden behind an event horizon is 
not globally naked, and such a spacetime would therefore satisfy 
the weak form of cosmic censorship. The strong form asks for more:
If an observer traverses the event horizon, will she encounter a
(locally) naked singularity? If so, physical predictions made on
the basis of the initial conditions will be upset by the presence 
of the singularity, and strong cosmic censorship will be violated.
Strong cosmic censorship therefore holds that no singularity
may be naked, even locally. 

The question asked in this essay is whether the laws of general 
relativity enforce strong cosmic censorship. 

\section{Black holes in asymptotically flat spacetime}
\label{poi-flat}

It is well known that the Reissner-Nordstr\"om, Kerr, and 
Kerr-Newman spacetimes contain timelike singularities inside their 
event horizons \cite{poi-4}. As these singularities are obviously naked, 
we must ask whether these spacetimes constitute a serious 
counter-example to strong cosmic censorship. I shall argue 
to the negative.

\begin{figure}
\begin{center}
\mbox{\epsfxsize=2.5in \epsffile{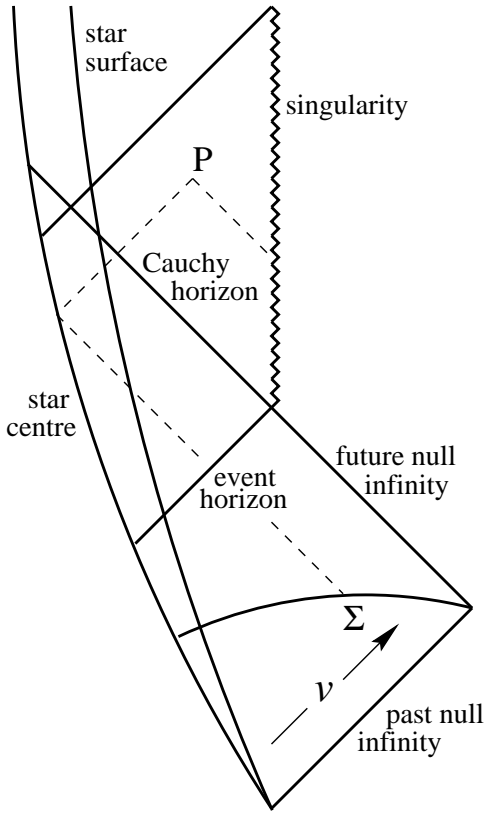}}
\end{center}
\caption{Conformal diagram of the Reissner-Nordstr\"om spacetime. 
The ingoing branch (left-going on the diagram) of the inner horizon
is a Cauchy horizon for the hypersurface $\Sigma$. This is illustrated 
by the fact that light rays originating from event P and propagating 
backward in time run into the timelike singularity at $r=0$. In this 
diagram, $v = \infty$ both at future null infinity and at the Cauchy 
horizon.}
\label{poi-fig2}
\end{figure}

Figure \ref{poi-fig2} shows a conformal diagram of the 
Reissner-Nordstr\"om  spacetime, whose metric is given by
\begin{equation}
ds^2 = -fdv^2 + 2dvdr + r^2 (d\theta^2 + \sin^2\theta\, d\phi^2),
\label{poi-1}
\end{equation}
where
\begin{equation}
f = 1 - \frac{2M}{r} + \frac{Q^2}{r^2}.
\label{poi-2}
\end{equation}
Here, $v$ is a null coordinate which is constant along radial
($d\theta=d\phi=0$), ingoing ($r$ decreasing) null geodesics;
$M$ denotes the mass of the black hole, and $Q$ its charge. The 
spacetime contains two types of horizons, located where $f=0$. 
The event horizon is at $r = r_e \equiv M + (M^2 - Q^2)^{1/2}$, 
while the inner horizon is at $r = r_i \equiv M - (M^2 - Q^2)^{1/2}$.
The diagram shows clearly that the ingoing branch of the inner
horizon is a Cauchy horizon for any spacelike hypersurface
$\Sigma$ preceding the formation of the event horizon. The
physical origin of the Cauchy horizon is also clear: predictions
made at any event P to the future of the Cauchy horizon would be 
upset by signals originating at the timelike singularity, where 
physics cannot be controlled. 

Clearly, the Reissner-Nordstr\"om spacetime is a counter-example
to strong cosmic censorship: the spacetime contains a Cauchy horizon, 
beyond which the evolution of physical fields becomes ambiguous. 
Furthermore, the same is true for the Kerr and Kerr-Newman spacetimes, 
which also contain Cauchy horizons. The real question, however, is 
whether these spacetimes constitute a {\it serious} counter-example 
to strong cosmic censorship. By this
I mean that if these spacetimes could be shown to form a set of
measure zero in some topological space of black-hole spacetimes, 
then they could be dismissed as inconsequential. On the other hand,
if black-hole spacetimes with Cauchy horizons formed an open set,
then we would have to conclude that strong cosmic censorship is
not enforced by general relativity.

The construction of such a topological space would be a difficult 
undertaking which, however, will be necessary to settle the issue. 
(This point was forcedly made to me by Jim Isenberg.) Here I will 
argue that the Kerr-Newman spacetimes should form a set of 
measure zero in any reasonable topological space of black-hole
spacetimes. 

Consider first the Reissner-Nordstr\"om spacetimes. (There is 
one spacetime for each value of the parameters $M$ and $Q$.) These 
spacetimes are very special, because they result from very special 
initial conditions: apart from being empty of any form of matter except 
for a static electric field, they are exactly spherically symmetric. 
However, it has long been known that slight deviations from 
these conditions, in the form of time-dependent matter fields or 
gravitational waves, produce large effects at the Cauchy 
horizon \cite{poi-5,poi-6,poi-7,poi-8,poi-9}. More precisely, 
physical quantities associated with the perturbations, such as 
the energy density measured by a free-falling observer, diverge 
at the Cauchy horizon. In other words, the Cauchy horizons of the 
Reissner-Nordstr\"om spacetimes are {\it unstable} to time-dependent 
perturbations. The same is true for the Kerr and Kerr-Newman spacetimes. 
This is an indication that these spacetimes must form a set of measure 
zero. 

This, however, is not good enough, because the perturbation analysis
involves only test fields in a fixed background spacetime. What must 
be understood, in a non-perturbative manner, is how the spacetime
itself evolves under the slightly different choice of initial 
conditions. This is the question that Werner Israel and I started
to examine in 1989 \cite{poi-10}. After a lot of work, carried out most 
notably by Israel's group in Canada and Amos Ori's group in Israel, 
the answer is now clear: Both for nonrotating and rotating black holes, 
the spacetime will develop a null curvature singularity at the
Cauchy horizon. This singularity is not of a big-crunch type as
in the Schwarzschild spacetime. Instead, it is characterized
by an infinite growth of the internal mass function at the Cauchy
horizon, whose area remains finite. This singularity is known as
the {\it mass-inflation} singularity. In effect, the perturbations
destroy the Cauchy horizon, and replace it with a null curvature
singularity. The mass-inflation scenario is now firmly established
in spherical symmetry, thanks to 
analytic \cite{poi-10,poi-11,poi-12,poi-13} and 
numerical \cite{poi-14,poi-15} calculations. The evidence 
is somewhat less firm, but still quite good, in the case of 
rotating black holes \cite{poi-16,poi-17,poi-18,poi-19}.

One might ask the following question: ``How is a null curvature
singularity any better than a Cauchy horizon? After all, isn't
predictability lost also at the singularity, because of the 
necessary breakdown of the classical laws there?'' Eanna 
Flanagan provided the following answer during the workshop,
which I fully endorse: The presence of a (nonsingular) Cauchy 
horizon inside a black hole is surprising because it
signals the breakdown of the classical laws without any {\it local} 
indication that something may be wrong. For example, a free-falling 
observer would measure the curvature tensor to be well below Planckian 
values, and would never suspect that classical general relativity was 
about to lose predictive power. This, clearly, is not the case near a 
curvature singularity. In a sense, the loss of predictability occurring 
at a Cauchy horizon is much worse than the ``mere'' breakdown of the 
classical laws near a curvature singularity. 

We may conclude that there is strong evidence that black-hole spacetimes 
with Cauchy horizons form a set of measure zero, because slight deviations 
in the initial conditions produce spacetimes whose causal structure is 
drastically different. In effect, slightly different initial conditions 
destroy the Cauchy horizon and replace it with a null curvature 
singularity. It is therefore tempting to suggest that the Kerr-Newman
spacetimes do not constitute a serious counter-example to 
strong cosmic censorship. In fact, I suspect that the following statement 
is true: In the topological space of all asymptotically-flat black-hole 
spacetimes, the set of all spacetimes containing a Cauchy horizon has
zero measure. Of course, no mathematically rigourous proof of this
statement exists.

\section{Cauchy-horizon instability}
\label{poi-inst}

It is useful to have a clear understanding of the physical
processes leading to the Cauchy-horizon instability. For
simplicity, we restrict attention to the Reissner-Nordstr\"om 
spacetime. 

We consider a simple model involving a test distribution of 
noninteracting massless particles. The particles originate 
from the region outside the black hole, move radially 
inward along curves of constant $v$, and eventually fall 
inside the black hole. They are described by the stress-energy 
tensor
\begin{equation}
T_{\alpha\beta} = \frac{L(v)}{4\pi r^2}\, 
(\partial_\alpha v) (\partial_\beta v),
\label{poi-3}
\end{equation}
where the luminosity function is given by $L(v) \sim v^{-p}$ 
(with $p$ a positive constant) when $v\to\infty$, to correctly 
reproduce Price's inverse-power law decay of radiative fields 
outside the black hole \cite{poi-20,poi-21,poi-22,poi-23}. 
We recall that $v=\infty$ designates both future null infinity 
and the Cauchy horizon (see Fig.~\ref{poi-fig2}).

The flux of particles is observed inside the black hole by a 
free-falling observer crossing the Cauchy horizon. This observer
moves on a radial geodesic, has a four-velocity $u^\alpha$, and
measures the energy density of the particles to be $\rho = 
T_{\alpha\beta} u^\alpha u^\beta = L(v) \dot{v}^2 /4\pi r^2$, 
where $\dot{v} \equiv u^v$. A simple calculation reveals that 
$\dot{v} \sim |\tilde{E}| \exp(\kappa_i v)$ when $v\to\infty$,
where $\tilde{E} \equiv - u_v$ is the observer's energy 
parameter, and 
\begin{equation}
\kappa_i = \frac{1}{2}\, \biggl| \frac{df}{dr} \biggr|_{r = r_i}
\label{poi-4}
\end{equation}
is the {\it surface gravity} of the inner horizon. Substitution 
yields
\begin{equation}
\rho \sim \frac{|\tilde{E}|^2}{4\pi {r_i}^2}\, 
v^{-p} e^{2\kappa_i v}, \qquad v \to \infty.
\label{poi-5}
\end{equation}
Thus, the measured energy density diverges as the observer
reaches the Cauchy horizon. In terms of $v$, this divergence is 
exponential; in terms of $\tau$, the amount of proper time left 
before reaching the Cauchy horizon, $\rho \sim 1/\tau^2$. The
physical interpretation is that the particles pile up at the 
Cauchy horizon, which is a surface of infinite blueshift. 
Consequently, the energy density diverge there. It is perhaps 
not surprising that a full backreaction calculation reveals 
the existence of a null curvature singularity at the Cauchy horizon. 

The extremal case ($|Q| = M$) requires a separate treatment, 
because $\kappa_i = 0$ for this spacetime. It turns out that 
in this case, the Cauchy horizon is {\it stable} to time-dependent 
perturbations \cite{poi-24}. The reason is that although an infinite 
blueshift still occurs at the Cauchy horizon, the blueshift factor 
diverges only as a power of $v$ instead of exponentially. Because 
$L(v)$ decays with a larger power, the energy density 
stays finite. This observation should not affect our conclusions, 
as presumably, extremal black holes also form a set of measure 
zero of spacetimes. 

\section{Black holes in non asymptotically flat spacetime}
\label{poi-deS}

While it appears highly plausible that strong cosmic censorship
is enforced for black holes residing in asymptotically flat 
spacetime, the same cannot be said for black holes residing in 
asymptotically de Sitter spacetime. 

\begin{figure}
\begin{center}
\mbox{\epsfxsize=3in \epsffile{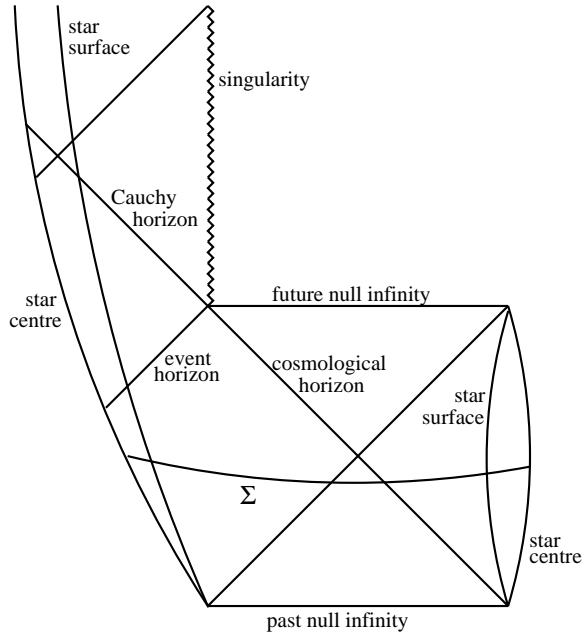}}
\end{center}
\caption{Conformal diagram of the Reissner-Nordstr\"om-de Sitter
spacetime. The ingoing branch of the inner horizon is a Cauchy
horizon for the hypersurface $\Sigma$. In this diagram, $v = \infty$ 
both at the cosmological horizon and at the Cauchy horizon.}
\label{poi-fig3}
\end{figure}

Consider first the Reissner-Nordstr\"om-de Sitter 
spacetime \cite{poi-25}, whose metric is given by 
Eq.~(\ref{poi-1}) with
\begin{equation}
f = 1 - \frac{2M}{r} + \frac{Q^2}{r^2} - 
\frac{1}{3} \Lambda r^2,
\label{poi-6}
\end{equation}
where $\Lambda$ is the cosmological constant, assumed to be 
positive. This spacetime contains three types of horizons,
situated at the three positive roots of $f$. The cosmological
horizon is located at $r=r_c$, the largest root. The black-hole
event horizon is at $r=r_e$, and the inner horizon at $r=r_i$ is
also a Cauchy horizon for any spacelike hypersurface $\Sigma$ lying 
outside the black hole. A conformal diagram of this spacetime is 
presented in Fig.~\ref{poi-fig3}. We see from the diagram that 
the cosmological horizon plays here the same role that
future null infinity played in the diagram of Fig.~\ref{poi-fig2}.

The major difference between this class of spacetimes and the
Reissner-Nordstr\"om class is that here, the parameters 
$\{M,Q,\Lambda\}$ can be chosen so that the Cauchy horizon is
{\it stable} to time-dependent perturbations. As was first shown 
by Brady and myself \cite{poi-1} on the basis of the simple argument 
presented in Sec.~\ref{poi-stab}, and then confirmed by Mellor
and Moss \cite{poi-26} on the basis of a complete perturbation analysis,
this happens whenever
\begin{equation}
\kappa_i \leq \kappa_c,
\label{poi-7}
\end{equation}
where $\kappa_i$ is the surface gravity of the inner horizon,
defined by Eq.~(\ref{poi-4}), and $\kappa_c = \frac{1}{2} 
|df/dr|(r_c)$ the surface gravity of the cosmological horizon.
Equation (\ref{poi-7}) defines a small but finite region of the 
parameter space $\{M,Q,\Lambda\}$. This region is described
in detail in Chambers' contribution to these proceedings. That 
the region must be small can be seen as follows (I thank Ian
Moss for providing me with this argument): In situations
close to realistic, the cosmological horizon would be at a very 
large radius, making $\kappa_c$ very small; to have $\kappa_i$ 
smaller than this requires the black hole to very nearly extremal, 
$|Q| = M(1-\epsilon)$, with $\epsilon$ a very small positive number. 

Equation (\ref{poi-7}) implies Cauchy-horizon stability not only
for the Reissner-Nordstr\"om-de Sitter spacetimes, but
also for the entire Kerr-Newman-de Sitter class. This was 
established by Chambers and Moss \cite{poi-27}, who also showed that 
Eq.~(\ref{poi-7}) defines a small but finite region of the 
parameter space $\{M,Q,a,\Lambda\}$, where $a$ is the black hole's
rotation parameter. Again, this is discussed in more detail in
Chambers' contribution. Equation (\ref{poi-7}) was also shown to 
imply stability in a fully nonperturbative analysis restricted to
spherical symmetry \cite{poi-28}.

If the open region of parameter space happens to correspond to
an open set in the topological space of black-hole spacetimes,
then we would be forced to conclude that strong cosmic censorship 
is {\it not} enforced by general relativity. There is, of course,
no rigourous proof that the spacetimes for which Eq.~(\ref{poi-7}) 
is satisfied do indeed form an open set, but I would regard this
as highly plausible. 

This example of an apparent violation of strong cosmic censorship
relies on the presence of a cosmological constant in the field 
equations, something which may be distasteful to some. However, 
another example was recently discovered by Horowitz and 
Sheinblatt \cite{poi-29}, and it does not require a cosmological constant. 
(I thank Gary Horowitz for pointing out this work to me.) These 
authors consider two oppositely charged black holes, each uniformly 
accelerating in a background magnetic field. The solution to the 
Einstein-Maxwell equations describing this spacetime is known as 
the Ernst solution \cite{poi-30}, and the causal structure of this 
spacetime is essentially identical to that of the 
Reissner-Nordstr\"om-de Sitter spacetime, except for the fact that 
the cosmological horizon is replaced by an acceleration horizon. 
Another similarity is that the spacetime is also not 
asymptotically flat. Horowitz and Sheinblatt show that the Cauchy 
horizon of the Ernst spacetime is {\it stable} whenever 
$\kappa_i \leq \kappa_a$, where $\kappa_a$ is the surface 
gravity of the acceleration horizon. Again, this 
inequality defines a finite region of parameter space, and it is 
tempting to suggest that this corresponds to an open set in the
topological space of black-hole spacetimes. If this were true, 
then strong cosmic censorship would be violated also for this class 
of spacetimes.

As we see, the fact that strong cosmic censorship might be violated 
for certain black-hole spacetimes is not necessarily due to the 
presence of a cosmological constant in the field equations. Instead, 
the essential features seem to be the presence of a ``large'' horizon 
and the absence of asymptotic flatness. I would conjecture the 
following statement: In the topological space of all black-hole 
spacetimes, including those which are not asymptotically flat, 
the set of all spacetimes containing a Cauchy horizon is open. If 
this statement is true, then strong cosmic censorship is not 
enforced by general relativity.

\section{Cauchy-horizon stability}
\label{poi-stab}

It is easy to understand why the Cauchy horizon of the 
Reissner-Nordstr\"om-de Sitter spacetime is stable when 
$\kappa_i \leq \kappa_c$. We consider once again the simple 
model of Sec.~\ref{poi-inst}, involving noninteracting 
massless particles described by the stress-energy of 
Eq.~(\ref{poi-3}). 

In Sec.~\ref{poi-inst}, the luminosity function $L(v)$ was taken 
to decay as an inverse-power law in order to correctly reproduce
the behaviour of radiative fields in the Reissner-Nordstr\"om
spacetime. This choice, however, is not appropriate for the
Reissner-Nordstr\"om-de Sitter spacetime \cite{poi-31}. To determine 
the correct behaviour of $L(v)$ we shall calculate $\rho_c$, the 
energy density of the infalling particles as measured by a 
free-falling energy crossing the cosmological horizon. 
We recall that $v=\infty$ designates
both the cosmological horizon and the Cauchy horizon 
(see Fig.~\ref{poi-fig3}). The steps are the same as in 
Sec.~\ref{poi-inst}, and we find
\begin{equation}
\rho_c \sim \frac{{\tilde{E}_c}^2}{4\pi {r_c}^2}\, 
L(v) e^{2\kappa_c v}, \qquad v \to \infty,
\label{poi-8}
\end{equation}
where $\tilde{E}_c$ is the observer's energy parameter. We demand
that the observer measure a finite, nonvanishing energy density as
she crosses the cosmological horizon. This requires $L(v)$ to decay
exponentially:
\begin{equation}
L(v) \sim K e^{-2\kappa_c v}, \qquad v \to \infty,
\label{poi-9}
\end{equation}
where $K$ is a constant. While Eq.~(\ref{poi-9}) serves mostly
to remedy the bad behaviour of the coordinates $v$ and $r$ at the
cosmological horizon, it also expresses the fact that this horizon 
is a surface of infinite redshift. 

Substituting Eq.~(\ref{poi-9}) into the calculation of 
Sec.~\ref{poi-inst} reveals that the energy density of the 
particles, as measured by an observer crossing the 
{\it Cauchy} horizon, is given by
\begin{equation}
\rho_i \sim \frac{|\tilde{E}_i|^2}{4\pi {r_i}^2}\, 
K e^{2(\kappa_i - \kappa_c) v}, \qquad v \to \infty.
\label{poi-10}
\end{equation}
We see that this quantity diverges when $\kappa_i > \kappa_c$,
but that it stays finite (in fact, goes to zero) when $\kappa_i
\leq \kappa_c$. This is just the condition expressed by
Eq.~(\ref{poi-7}). The interpretation is clear. The factor
$\exp(2\kappa_i v)$ comes from the infinite blueshift occurring 
at the inner horizon, while the factor $\exp(-2\kappa_c v)$ comes 
from the infinite redshift occurring at the cosmological horizon. 
When $\kappa_i > \kappa_c$ the blueshift wins over the redshift,
and the Cauchy horizon is unstable. On the other hand, when
$\kappa_i \leq \kappa_c$ the redshift wins, and the Cauchy
horizon is stable.

\section{Quantum effects}
\label{poi-quantum}

We now consider the quantum stability of the
Reissner-Nordstr\"om-de Sitter spacetime. This question is 
addressed by admitting the existence of quantized matter 
fields in the spacetime, and examining the behaviour of 
$\langle T^{\alpha\beta} \rangle$, the renormalized 
expectation value of their stress-energy tensor, near
the Cauchy horizon. This is a difficult question, even
when no attempt is made to take into account the spacetime's
response to the quantum stress-energy tensor. The difficulty 
resides in the calculation of $\langle T^{\alpha\beta} \rangle$, 
even in a fixed spacetime possessing spherical symmetry. The
difficulty is even more acute in our case, because (as will 
become clear below) the quantum state cannot be chosen among 
the standard ones, such as the Hartle-Hawking or Unruh vacua. 

We shall therefore consider a simpler problem, that of 
quantizing matter fields in a two-dimensional version 
of the Reissner-Nordstr\"om-de Sitter spacetime, with 
metric
\begin{equation}
ds^2 = -f\, du dv,
\label{poi-11}
\end{equation}
where $f$ is given by Eq.~(\ref{poi-6}) and the null coordinate 
$u$ is defined by $du = dv-2f^{-1}dr$. (It should be noted that 
the coordinates $u$ and $v$ cannot be extended across the event 
horizon. This will not be a problem for this calculation.) In 
two dimensions, $\langle T^{ab} \rangle$ can be computed 
explicitly \cite{poi-32}; we will do so for a conformally invariant 
scalar field. 

The calculation of $\langle T^{ab} \rangle$ starts with the 
trace-anomaly equation \cite{poi-33},
\begin{equation}
\langle T \rangle = \alpha R,
\label{poi-12}
\end{equation}
where $\alpha = (24\pi)^{-1}$ and $R = -f''$ is the Ricci scalar
associated with the metric of Eq.~(\ref{poi-11}); primes indicate
differentiation with respect to $r$. This equation immediately
gives us one component of the stress-energy tensor,
\begin{equation}
\langle T_{uv} \rangle = \frac{\alpha}{4}\, f f''.
\label{poi-13}
\end{equation}
The others are obtained by integrating the conservation
equations, $\langle T^{ab} \rangle_{;b} = 0$. A straightforward 
calculation yields 
\begin{equation}
\langle T_{uu} \rangle = 
- \frac{\alpha}{2}\, \Bigl[ F(r) - A(u) \Bigr]
\label{poi-14}
\end{equation}
and
\begin{equation}
\langle T_{vv} \rangle = 
- \frac{\alpha}{2}\, \Bigl[ F(r) - B(v) \Bigr],
\label{poi-15}
\end{equation}
where
\begin{equation}
F(r) = \frac{1}{4}\, \Bigl( f'^2 - 2 f f'' \Bigr),
\label{poi-16}
\end{equation}
while $A(u)$ and $B(v)$ are arbitrary functions which serve to define 
the state of the quantum field. We demand that this state be regular on
the cosmological and event horizons, so that the quantum stress-energy
tensor will also be regular there.  

To see what requirements must be made on $A(u)$ and $B(v)$, we construct 
$\langle \rho \rangle \equiv \langle T_{ab} \rangle u^a u^b$, the energy 
density as measured by a free-falling observer with two-velocity $u^a$. 
This has components 
\begin{equation}
u^v = \frac{\tilde{E} \pm (\tilde{E}^2 - f)^{1/2}}{f}
\label{poi-17}
\end{equation}
and
\begin{equation}
u^u = \frac{\tilde{E} \mp (\tilde{E}^2 - f)^{1/2}}{f},
\label{poi-18}
\end{equation}
where $\tilde{E}$ is the observer's energy parameter, and
the upper (lower) sign is chosen if $r$ is increasing (decreasing) 
along the world line. We obtain
\begin{eqnarray}
\langle \rho \rangle &=& 
-\frac{\alpha}{2f^2}\, \Bigl[ (2 \tilde{E}^2 - f)(2F-A-B)
\nonumber \\ & & \mbox{}
\pm 2 \tilde{E} (\tilde{E}^2 - f)^{1/2} (A-B) - f^2 f'' \Bigr].
\label{poi-19}
\end{eqnarray}
We now want to evaluate $\langle \rho \rangle$ near the horizon
$r=r_j$, where $r_j$ stands for either $r_c$ (cosmological horizon), 
$r_e$ (event horizon), or $r_i$ (Cauchy horizon). A straightforward 
calculation shows that near $f=0$,
Eq.~(\ref{poi-19}) reduces to
\begin{eqnarray}
\langle \rho \rangle &=& -\frac{\alpha}{2f^2}\, \Bigl\{ 
\bigl[ 2{\kappa_j}^2 - (1 \mp \epsilon) A - (1\pm \epsilon) B \bigr]
\nonumber \\ & & \mbox{} \times
(2 \tilde{E}^2 - f) + O(f^2) \Bigr\},
\label{poi-20}
\end{eqnarray}
where $\kappa_j = \frac{1}{2}|f'(r_j)|$ is the surface gravity
of the horizon under consideration, and $\epsilon \equiv 
\mbox{sign}(\tilde{E})$. Notice that we have never used the
detailed form of the function $f(r)$ in this calculation. 

We first evaluate $\langle \rho \rangle$ at the cosmological
horizon, where $r=r_c$ and $v=\infty$. An observer crossing 
this horizon has a positive energy parameter ($\epsilon = +1$)
and moves in the direction of increasing $r$ (upper sign). The 
term within the square brackets therefore reduces to 
$2{\kappa_c}^2 - 2 B(\infty)$. If we demand
\begin{equation}
B(v\to\infty) = {\kappa_c}^2 + O(e^{-2\kappa_c v}),
\label{poi-21}
\end{equation}
then this term will be $O(f^2)$ and $\langle \rho \rangle$
will be well-behaved at the cosmological horizon. Thus, our
choice of quantum state is restricted by Eq.~(\ref{poi-21}).

Next, we evaluate $\langle \rho \rangle$ at the event horizon,
where $r=r_e$ and $u=\infty$. An observer crossing this horizon
has a positive energy parameter ($\epsilon = +1$) and moves
in the direction of decreasing $r$ (lower sign). The square-brackets 
term becomes $2{\kappa_e}^2 - 2 A(\infty)$. If we demand
\begin{equation}
A(u\to\infty) = {\kappa_e}^2 + O(e^{-2\kappa_e u}),
\label{poi-22}
\end{equation}
then this term will be $O(f^2)$ and $\langle \rho \rangle$ will 
be well-behaved at the event horizon. Thus, our choice of 
quantum state is further restricted by Eq.~(\ref{poi-22}).
It is interesting to note that only the asymptotic behaviours of
$A(u)$ and $B(v)$ are restricted by the regularity conditions.
Otherwise, these functions are completely arbitrary, allowing for
much freedom in the choice of the quantum state. A {\it particular} 
quantum state meeting the regularity requirements is the 
Markovi\'c-Unruh vacuum \cite{poi-34}.

Finally, we evaluate $\langle \rho \rangle$ at the Cauchy horizon,
where $r=r_i$ and $v=\infty$. Since an observer crossing this horizon
moves with a negative energy parameter in the direction of decreasing
$r$, we must use $\epsilon = -1$ and the lower sign. The square-brackets 
term becomes $2 {\kappa_i}^2 - 2 B(\infty)$ which, by virtue of
Eq.~(\ref{poi-21}), is $2({\kappa_i}^2 - {\kappa_c}^2)$. This
gives 
\begin{equation}
\langle \rho \rangle \sim 2 \alpha \tilde{E}^2 
({\kappa_i}^2 - {\kappa_c}^2)\, \frac{1}{f^2}.
\label{poi-23}
\end{equation}
Thus, $\langle \rho \rangle$ diverges at the Cauchy horizon.

We have proved the following theorem:
\begin{verse}
In two-dimensional Reissner-Nordstr\"om-de Sitter spacetime, for 
{\it any} quantum state regular on both the cosmological horizon
and the event horizon, the renormalized expectation value of the
stress-energy tensor of a conformally invariant scalar field 
diverges at the Cauchy horizon, except when $\kappa_i = \kappa_c$.
\end{verse}
The theorem implies that the two-dimensional spacetime is
quantum mechanically unstable, except for the set of measure
zero of spacetimes for which $\kappa_i = \kappa_c$. A similar
theorem is believed to hold also for the Ernst spacetime \cite{poi-29}.

What does this theorem tell us about the four-dimensional world?
Although we shall not go into these details here \cite{poi-2}, physical 
intuition suggests that the four-dimensional spacetime must also be 
quantum mechanically unstable. Indeed, it appears that the quantum 
physics of the Cauchy-horizon instability, which is revealed by the
two-dimensional calculation, is robust and does not depend
on the dimensionality of spacetime nor the nature of the 
quantum field (its spin, conformal invariance, etc.). The
quantum mechanical instability can be intuitively explained
in terms of fundamental processes such as the creation of
thermal quanta near horizons, and the gravitational redshifts
and blueshifts that these quanta undergo. Since these processes
take place equally well in four as in two dimensions, I am
quite confident that the quantum stress-energy tensor will diverge
also at the Cauchy horizon of the four-dimensional spacetime.

\section{Conclusion}
\label{poi-conclusion}

I will conclude with these two statements:
\begin{itemize}
\item Black-hole Cauchy horizons may be classically stable if the
black hole does not reside in asymptotically flat spacetime. This
suggests that strong cosmic censorship is not enforced by the
classical formulation of general relativity.
\item Black-hole Cauchy horizons are always quantum mechanically
unstable, except possibly for a set of measure zero of spacetimes.
This suggests that strong cosmic censorship might be enforced by
the semi-classical formulation of general relativity.
\end{itemize}
Although these statements are clearly not supported by rigourous
mathematical proofs, I regard the evidence for their validity as 
quite compelling. (Of course, the second part of the second 
statement is pure speculation.) 

To the extent that these statements are true, it is most intriguing 
that quantum physics must be invoked in order to restore the full
predictive power of general relativity. Here we may notice an 
interesting similarity with the physics of chronology 
horizons \cite{poi-35}. 

And to the extent that the first statement is true, the following
question remains: ``Does the semi-classical formulation of general
relativity really enforce strong cosmic censorship?''

\section*{Acknowledgments}

This work was supported by the Natural Sciences and Engineering
Research Council of Canada. It is a pleasure to thank Lior Burko,
Amos Ori, and Liz Youdim for having organized such a wonderful 
workshop.

\end{document}